# Threshold Voltage variation with respect to Gate geometry in Nano-scale MOSFETS

Raja N. Mir

*Abstract*— The tremendous progress in Metal Oxide Semiconductor (MOS) technology has been a direct consequence of device scaling for past several decades. But as we have entered the nanometer era many problems related to leakage currents and other issues related to variability impacting the yield are of concern. Herein we have investigated how the change in the Fin Architecture and Gate Length of the MOS device impacts the Threshold Voltage.

I. INTRODUCTION

With ever increasing digital performance and application, the demands are very high for better and higher performance digital semiconductor devices. This is directly dependent on the semiconductor industry being able to scale the Metal Oxide Semiconductor Field Effect Transistor (MOSFET). With many innovations in the gate-stack technology, we are able to push the physical gate length 14 nm and this will go well beyond this scale. The High K metal gate reduces the gate leakage significantly by improving the electrostatics within the MOSFET channel. Another way to improve the electrostatic situation in the MOSFET channel is to optimize the MOSFET channel structure or geometry. The further scaling will continue to face increasing number of technological and physical challenges, among these the structural optimization of the channel is also one of the major concerns. The other challenges being able to control the variability in doping, poly-silicon granularity, random grain orientation within the High-K dielectric [1-3].

In the present work we have full three-dimensional simulation of the gate-structure impact on the nano-scale MOSFET threshold characteristics. We have also done some experimental studies of this phenomena. In addition, there are full-bandstructure tight binding calculations which can provide further insight but are not discussed in this paper [4-13]. In the next section we describe the simulation setup, and results of the gate structure changes on the transistor threshold voltage ( Vt ).

II. SIMULATION SETUP

TCAD based tools[14] are widely used to simulate electrical characteristics of MOS transistors, e.g:- Vth, SS, Ion, Ioff , gm, gds etc. These characteristics are very important metrics of device behavior and are used to compare and contrast different devices. Within the framework of TCAD they are determined by first doing the process simulation then doing a device simulation and finally an extraction from the the results of the device simulation. We compared the behavior of four devices: planar MOSFET , triangular bulk 3D-MOSFET, rectangular bulk 3D-MOSFET and rectangular SOI 3D-MOSFET. These analyses will indicate the superiority of the 3D-MOSFET over the planar device. The 3D-MOSFET height is equal to 24 nm, the base width is 10.7 nm. The angle of the fin in the triangular 3D-MOSFET is equal to 84 degrees. These devices have the same underlying process. The planar device is obtained by modifying the 3D-MOSFET, by reducing the fin height(H) and stretching the fin width(W) equal to the $(2*H + W)$. The doping concentration in the source and the drain is $10^{20}$ cm3 and that in the channel region is $10^{18}$ cm3. We have closely tried to follow Intel Ivy Bridge [1] device dimensions.

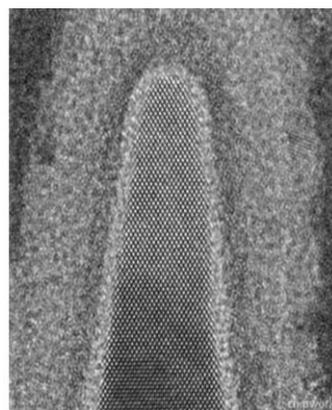

Figure 1: TEM Lattice image of NMOS fin structure. Adapted from [15]

III. RESULTS

In the Figure:2 we see that the threshold voltage of the 3D-MOSFET's is very close as we move down the gate length. We observe a voltage rollon trend, which is there because of the fact that the halo implants overlap as the source and the drain come closer together. In case of the planar device the threshold voltage is more than 3D-MOSFET's. Overall a voltage rolloff in case of the planar device is observed but the rolloff is not very clear as there are two competing effects going on in the MOSFET. One is that the Drain Induced Barrier Lowering ( DIBL ) effect will tend to give a rolloff but on the other hand due to source and drain halo implants the voltage rollon should take place. The combined result of these two effects in this specific case resulted in overall lowering of the threshold voltage as we move down the gate length.

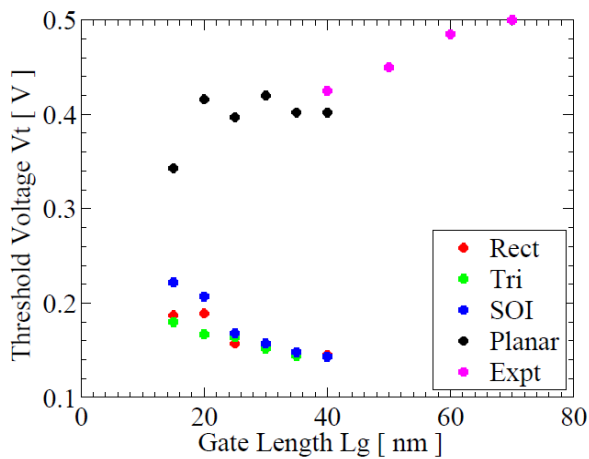

**Figure 2: Threshold voltage for various devices with similar gate lengths and widths.**

IV. CONCLUSION

It is clearly seen that the threshold voltage of the 3D-MOSFET is better than the planar counterpart. The voltage roll-on for the 3D MOSFET is evident as the channel is reduced, the roll-off for the planar counter part is also observed to some degree.